\documentstyle[11pt,psfig,newpasp,twoside]{article}
\index{neutron stars!cooling}

\def\edcomment#1{\iffalse\marginpar{\raggedright\sl#1\/}\else\relax\fi}
\reversemarginpar

\begin{document}
\title{Anisotropic Heat Transfer inside Rotating Neutron Stars}
 \author{C. Y. Hui}
\affil{Department of Physics, The University of Hong Kong, Hong Kong, China} 
\author{K. S. Cheng}
\affil{Department of Physics, The University of Hong Kong, Hong Kong, China} 

\begin{abstract}
We have developed the anisotropic heat transport equation for rotating neutron stars. 
With a simple model of neutron star, we also model the propagation of heat pulses result from transient energy releases inside the star. 
Even in slow rotation limit, 
the results with rotational effects involved could differ significantly from those obtained with a spherically symmetric metric in the time scale of the thermal afterglow.
\end{abstract}

\section{Introduction}
In our study, we have investigated the effects of rotation on the thermal afterglows result from transient energy releases inside the star. Three possible forms of energy release, namely, 
the `shell', `ring' and `spot' cases, are considered. `Shell' case is a hypothetical case that energy is released at a particular density in the form of a spherical 
shell. In `ring' case, energy is released at a particular density around the rotational equator. This could be resulted from superfluid-driven glitch (Anderson 1975). 
In `spot' case, energy is released at a localized region. This could be resulted from crust-cracking (Ruderman 1969).

\section{Model}
The general expression for the line element of an axial symmetric spacetime is determined by time-translational invariance and 
axial-rotational invariance:
\begin{equation}
ds^{2}=e^{2\nu \left( r,\theta \right) }dt^{2}-e^{2\lambda \left(
r,\theta \right) }dr^{2}-r^{2}e^{2\psi \left( r,\theta \right) }\left[
d\theta ^{2}+\sin ^{2}\theta \left( d\phi -\omega \left( r,\theta \right)
dt\right) ^{2}\right]
\end{equation}
We calculate all the necessary metric functions as well as the stellar structure by using Hartle's formalism (Hartle \& Thorne 1968).\\
The energy momentum tensor inside a star consist of perfect fluid which allow heat flow is:
\begin{equation}
T^{\mu\nu}=\left( \rho+P\right) u^{\mu}u^{\nu}-Pg^{\mu\nu}+u^{\mu}q^{\nu}+u^{\nu}q^{\mu}
\end{equation}
With the rotating metric, the vanishing covariant divergence of the energy momentum tensor gives us the heat transport equation.
\begin{equation}
T^{t\alpha}_{;\alpha}=T^{tt}_{;t}+T^{tr}_{;r}+T^{t\theta}_{;\theta}+T^{t\phi}_{;\phi}=0
\end{equation}
The equation is checked by recovering to the non-rotational case (Cheng, Li, \& Suen 1998) and 
the Newtonian case with spherically symmetric metric and flat spacetime respectively.

\section{Monte Carlo Simulations}
The equation of state of neutron matter 
(Pandharipande 1971) is adopted. We also adopt the heat capacities, thermal conductivity and neutrino emissivities from previous studies (Maxwell 1979;  
Flowers \& Itoh 1981).\\ 
We simulate the propagation of heat pulses by Markovian random walk. Models with different rotational frequency are compared.

\begin{figure}[h]
\centerline{
\psfig{file=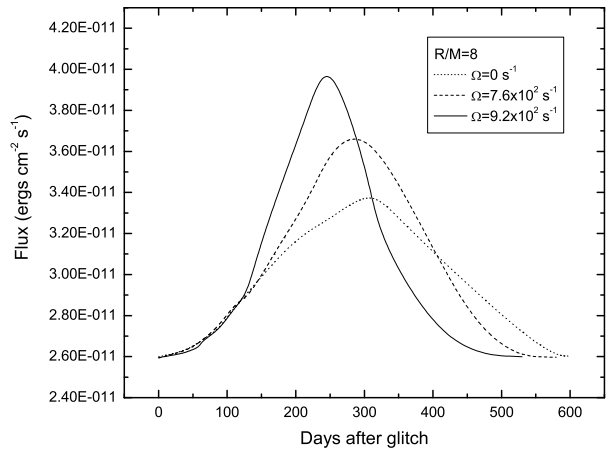,width=3.5in}
            }
\caption{The evolution curves of the flux of thermal X-ray for a neutron star. Three cases with different rotational frequency are compared.}
\end{figure}

\end{document}